\documentclass[useAMS,usenatbib]{mn2e}
\usepackage{graphicx}
\usepackage{natbib} 

\title[Spin alignment of galaxies in filaments]{Evidence for spin alignment of spiral and elliptical/S0 galaxies in filaments}
\author[E. Tempel, R.~S. Stoica and E. Saar]{E. Tempel$^{1,2}$\thanks{E-mail:
elmo@aai.ee}, R.~S. Stoica$^{3,4}$ and E. Saar$^{1,5}$\\
$^{1}$Tartu Observatory, Observatooriumi~1, 61602 T\~oravere, Estonia\\
$^{2}$National Institute of Chemical Physics and Biophysics, R\"avala pst 10, Tallinn 10143, Estonia\\
$^{3}$Universit\'e Lille 1, Laboratoire Paul Painlev\'e, 59655 Villeneuve d'Ascq Cedex, France\\
$^{4}$Institut de M\'ecanique C\'eleste et Calcul d'Eph\'em\'erides, Observatoire de Paris, 75014 Paris, France\\
$^{5}$Estonian Academy of Sciences, Kohtu 6, Tallinn 10130, Estonia}

\begin{document}

\date{Accepted 2012 October 5.  Received 2012 October 5; in original form 2012 June 30}

\pagerange{\pageref{firstpage}--\pageref{lastpage}} \pubyear{2012}

\maketitle

\label{firstpage}

\begin{abstract}
Galaxies are not distributed randomly in the cosmic web but are instead arranged in filaments and sheets surrounding cosmic voids. Observationally there is still no convincing evidence of a link between the properties of galaxies and their host structures. However, by the tidal torque theory (our understanding of the origin of galaxy angular momentum), such a link should exist. Using the presently largest spectroscopic galaxy redshift survey (SDSS) we study the connection between the spin axes of galaxies and the orientation of their host filaments.

We use a three dimensional field of orientations to describe cosmic filaments. To restore the inclination angles of galaxies, we use a 3D photometric model of galaxies that gives these angles more accurately than traditional 2D models.

We found evidence that the spin axes of bright spiral galaxies have a weak tendency to be aligned parallel to filaments. For elliptical/S0 galaxies, we have a statistically significant result that their spin axes are aligned preferentially perpendicular to the host filaments; we show that this signal practically does not depend on the accuracy of the estimated inclination angles for elliptical/S0 galaxies.

\end{abstract}


\section{Introduction}

Disc-dominated galaxies are the morphologically dominant class of galaxies in the present-day Universe \citep{Bamford:09}. As these galaxies are rotationally supported, it is of vital importance to understand how disc galaxies acquire their angular momentum. The tidal-torque theory predicts alignment effects of disc galaxies, since the acquisition of the angular momentum  is partially governed by environmental effects, such as tidal shearing produced by the neighbouring primordial matter distribution and the moment of inertia of the forming protogalaxy \citep[for a review see][]{Schafer:09}. Hence, testing the alignment of the orientation of galaxies with that of their surrounding environment (e.g., filaments) provides vital information about how galaxies form in the cosmic web.

It is known that angular momenta of neighbour galaxies are correlated \citep{Slosar:09, Lee:11, Andrae:11}, indicating that the environment influences the acquisition of the angular momentum. However, \citet{Andrae:11} argue that this correlation is plausible but not statistically significant.

During the last years, correlations between dark matter haloes and their host filaments have been detected in $N$-body simulations \citep{Hatton:01, Faltenbacher:02, Bailin:05, Altay:06, Brunino:07, Wang:11}. More specifically, \citet{Aragon-Calvo:07}, \citet{Trowland:12}, and \citet{Codis:12} showed that the orientation of the halo spin vector is mass-dependent. Spin axes of low-mass haloes tend to be aligned parallel to the filaments, whereas high mass haloes have an orthogonal alignment. In this picture, low-mass haloes form through the winding of flows in cosmic sheets/walls; hence at the intersection of these walls (in filaments), the spin axes of the haloes are parallel to the filaments. On the other hand, massive haloes are typically the product of mergers along such a filamentary structure, and thus their spin axes are perpendicular to the spine of the filaments.

Previous studies based on observed galaxy catalogues have also indicated that spin axes of galaxies tend to be correlated with the structures in which they are embedded \citep{Kashikawa:92, Navarro:04a, Trujillo:06}. These studies concentrate on sheets and voids and show that the spin axes of galaxies are preferentially oriented perpendicular to the host structures. In a more recent study, \citet{Jones:10} used a small sample of edge-on galaxies from the SDSS and showed that the spin axes of these galaxies are aligned perpendicular to the spine of the parent filament. These objects are also less bright, and so their results are in contradiction with the results of $N$-body simulations.

The picture is even more complicated since the spin axes of galaxies and their host haloes are not strictly parallel; \citet{van-den-Bosch:02} find a median misalignment of approximately $30\degr$. Furthermore, minor mergers can significantly disturb the angular momenta of galaxies \citep{Moster:10}.

To add even more controversy to the picture,  \citet{Slosar:09a} claimed recently that in contrast to previous studies, there are no correlations between galaxies and the voids in which they are located. Furthermore, \citet{Varela:12} used the SDSS data and the morphological classification from the Galaxy Zoo project and found that galaxy spins tend to be perpendicular to the void walls they are located in, which is in contradiction with the results of \citet{Navarro:04a} and \citet{Trujillo:06}. Thus, the results remain contradictory, mostly  because of the difficulty to measure the inclination angles of galaxies and/or to properly define the large-scale filamentary structures.

In the present study, we use the full SDSS spectroscopic sample to study the correlation between the spin axes of galaxies and the orientation of their host filaments. We build a field of orientations for the filaments. We use a three-dimensional galaxy model to estimate the inclination angles of galaxies. To have good estimates of the inclination angles, we limit our sample strictly to galaxies with the bulge-to-disc ratio less than three: e.g. spirals and elliptical/S0 galaxies.

The structure of the present paper is as follows: in Sect.~\ref{sect:data} we briefly describe the used data. In Sect.~\ref{sect:model} we describe the three-dimensional galaxy model and explain how we selected the galaxy sample for the correlation study. The algorithm to build the filament orientation field is described in Sect.~\ref{sect:fil} and the correlation estimator is defined in Sect.~\ref{sect:cor}. Finally, in Sect.~\ref{sect:results} we give the results and discuss them. 

Throughout this paper we assume the following cosmology: the Hubble constant $H_0 = 100\,h\,\mathrm{km\,s^{-1}Mpc^{-1}}$, the matter density $\Omega_\mathrm{m}=0.27$ and the dark energy density $\Omega_\Lambda=0.73$ \citep{Komatsu:11}.

\section{SDSS data}
\label{sect:data}

Our present study is based on the SDSS DR8 \citep{Aihara:11}. We use only the main contiguous area of the survey (the Legacy Survey) and the spectroscopic galaxy sample as compiled in \citet{Tempel:12}. The lower Petrosian magnitude limit for this sample is set to $m_r=17.77$, since for fainter galaxies, the spectroscopic sample is incomplete \citep{Strauss:02}. To exclude the Local Supercluster from the sample, the lower CMB-corrected distance limit $z=0.009$ was used. The upper limit was set to $z=0.2$, since at larger distances the sample becomes very diluted. The sample includes 576493 galaxies.

The DR8 sample is greatly improved over the previous data releases, in the sense that all the imaging data have been reprocessed using an improved sky-subtraction algorithm, a self-consistent photometric recalibration, and the flat-field determination. This improvement makes the DR8 a good sample for detailed photometric modelling.

For photometric modelling we used the atlas (prefix `fpAtlas'), mask (prefix `fM') and point-spread-function (prefix `psField') images of the sample galaxies. The images were downloaded from the SDSS Data Archive Server (DAS), and the SDSS software utilities \texttt{readAtlasImages} and \texttt{readPSF} were used for the atlas and PSF images, respectively. Masks were applied to the atlas images, using galaxy positions (given by \texttt{rowc} and \texttt{colc} in SDSS CAS).

For photometric modelling, the initial guess for the centre, position, and size of a galaxy is taken from the SDSS Catalog Archive Server (CAS), where each galaxy has been approximated by simple de~Vaucouleurs and exponential models. To model the structural parameters, only the $gri$ filters were used, since the uncertainties for the $u$ and $z$ imaging are the largest.

To find the filamentary structure in the SDSS data, we have to suppress first the finger-of-god redshift distortions for groups. For that we use the Friends-of-Friends (FoF) groups compiled in \citet{Tempel:12}: the details of the group finding algorithm are explained in \citet{Tago:08, Tago:10}.
We spherisize the groups using the rms sizes of galaxy groups in the sky and their rms radial velocities as described in \citet{Liivamagi:12} and \citet{Tempel:12}. Such a compression will lead to a better estimate of the density field and can  help to find the real filamentary structure.

To define the filamentary structure, we use a Cartesian grid based on the SDSS angular coordinates $\eta$ and $\lambda$, allowing the most efficient placing of the galaxy sample cone inside a box. The galaxy coordinates are calculated as follows:
\begin{eqnarray}
    x&=&-d_\mathrm{gal}\sin{\lambda}\nonumber\\
    y&=&d_\mathrm{gal}\cos{\lambda}\cos{\eta}\\ \label{eq:1}
    z&=&d_\mathrm{gal}\cos{\lambda}\sin{\eta}\nonumber ,
\end{eqnarray}
where $d_\mathrm{gal}$ is the finger-of-god suppressed co-moving distance to a galaxy.

The main purpose of this paper is to study the correlation between the spin vectors of galaxies and the spine/axis of filaments. Since the full angular momentum of galaxies cannot be estimated from photometric observations, we use the normalised spin vectors that describe the rotation axis of galaxy. Moreover, the spin vectors of galaxies (specifically, the inclination angles of galaxies) can in principle estimated only for disc-dominated galaxies: for pure elliptical galaxies the inclination angle cannot be found uniquely from photometric observations. To select spiral galaxies and lenticular (S0) galaxies, we use the morphological classification as described in \citet{Tempel:11a}, where we divide galaxies into three classes: spirals, ellipticals, and with uncertain classification. This classification takes into account the SDSS model fits, apparent ellipticities (and apparent sizes), and different galaxy colours. In \citet{Tempel:12} we compared this classification with the classification by the Galaxy Zoo project \citep{Huertas-Company:11} and showed that they are in very good agreement.

So, for finding the filaments and their orientation we use all galaxies in the subsample of the SDSS as selected in \citet{Tempel:12}. For correlation studies we restrict our sample to the galaxies with the bulge-to-disc ratio less than three (see Sect.~\ref{sect:galsample}).

\section[]{Three-dimensional photometric modelling} 
\label{sect:model}

In this section we describe the 3D photometric model and describe in detail the sample of galaxies that we will use to study the correlations between galaxies and filaments. Full results of the photometric modelling of the SDSS main galaxy sample together with a catalogue will be published in a separate paper.

\subsection{Description of the three-dimensional photometric model of galaxies}

We use a three-dimensional model to describe the structure of a galaxy. The model is fully described in \citet{Tempel:10, Tempel:11} (in the case of M\,31). In \citet{Tempel:12a} we applied the model to a test sample of SDSS galaxies and showed that it can be used to automatically model the SDSS galaxies with a sufficiently good accuracy. Especially important for the present study is that the inclination angles of galaxies can be restored very accurately: for the majority of disc-dominated galaxies the differences are less than 5 degrees; for galaxies with the bulge-to-disc ratio less than three the accuracy is better than 10--15 degrees.

For consistency reasons we give a short description of the model below.

The model galaxy is given as a superposition of its individual stellar components. In the present study we use a two component model: a spheroid\,+\,a disc. Both components are approximated by an ellipsoid of revolution with a constant axial ratio $q$; its spatial density distribution follows Einasto's law
\begin{equation}
    l(a)=l(0)\exp{\left[-\left(\frac{a}{ka_0}\right)^{1/N}\right]},
\end{equation}
where $l(0)=hL/(4\upi qa_0^3)$ is the central density and $L$ is the luminosity of the component; $a=\sqrt{r^2+z^2/q^2}$, where $r$ and $z$ are the cylindrical coordinates. We use  the harmonic mean radius $a_0$ to describe the extent of the component; $h$ and $k$ are normalising parameters depending on the structural parameter $N$ \citep[see appendix~B of][]{Tamm:12}.

The density distributions of both components are projected along the line-of-sight and their sum yields the surface brightness distribution of the model
\begin{equation}
    \label{eq:los_dens}
    f_\mathrm{model}(x,y)=2\sum\limits_{j=1}^2 \frac{q_j}{Q_j}\int\limits_{A_j}^\infty \frac{l_j(a)a\mathrm{d}a}{\left(a^2-A_j^2\right)^{1/2}},
\end{equation}
where $A_j^2=x^2+y^2/Q_j^2$ is the major semi-axis of the equidensity ellipse of the projected light distribution and $Q_j$ are the apparent axial ratios of the components ($Q^2=\cos^2{i}+q^2\sin^2{i}$); $i$ is the inclination angle between the spin vector of the galaxy and the line-of-sight. The summation is over the two visible components (the spheroid and the disc).
Equation~(\ref{eq:los_dens}) gives the surface brightness of the galaxy for all lines-of-sight that were used to create the model image of a galaxy. Before comparing model images with observations we convolved the model images with the SDSS point spread function (PSF) for a given galaxy.

The correctness of the model fit was estimated by using the $\chi^2$ value, defined by
\begin{equation}
  \chi^2 = \frac{1}{N_\mathrm{dof}}\sum\limits_{\nu}\sum\limits_{x,y\in \mathrm{mask}}
  \frac{\left[f_\mathrm{obs}^\nu(x,y)-f_\mathrm{model}^\nu(x,y)\right]^2}{\sigma^\nu(x,y)^2},
  \label{eq:chi}
\end{equation}
where $N_\mathrm{dof}$ is the number of degrees of freedom in the fit; $f_\mathrm{obs}^\nu(x,y)$ and $f_\mathrm{model}^\nu(x,y)$ are the observed and modelled fluxes at the given pixel $(x,y)$ and index $\nu$ indicates the filter ($gri$); $\sigma(x,y)$ is the Poisson error at each pixel \citep{Howell:06}. The summation is taken over all filters $(\nu)$ and all pixels of each galaxy as defined by the corresponding mask.

To minimise the $\chi^2$ we have used the downhill simplex method of Nelder and Mead from the Numerical Recipe library \citep{Press:92}. This method is quite simple and efficient in searching large parameter spaces. The test sample study in \citet{Tempel:12a} showed that the method performs well and gives sufficiently accurate results with a reasonable time. In \citet{Tempel:12a} we also gave an example how this model performs for the real SDSS data; there are no systematic trends and the model performs well.

\subsection{Estimation of the inclination angles of galaxies}

\begin{figure}
    \includegraphics[width=84mm]{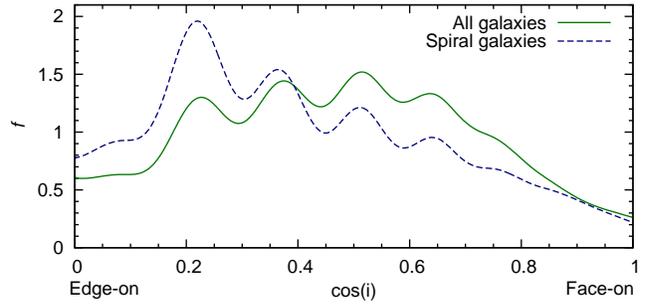}
    \caption{Distribution of the inclination angles of galaxies. The green solid line shows the distribution for all galaxies; the blue dashed line shows the distribution for spiral galaxies. Spiral galaxies were selected based on the classification given in \citet{Tempel:11a}. The peaks in the distribution are the remnant of the modelling procedure (see text for more information).}
     \label{fig:cosidist}
\end{figure}

Since our method to model galaxies depends slightly on the initial parameters, due to the degeneracy between various parameters, we cannot find all the parameters directly. The degeneracy is strongest for the inclination angles of galaxies, which are strongly degenerate with the thickness/ellipticity of the galaxy.

To measure the inclination angle of a galaxy, we run the modelling code with ten different initial guesses for the inclination angle between $0\degr$ (face-on galaxies) and $90\degr$ (edge-on galaxies). This modelling was done for the $gri$ filters separately, in order to have three different guesses for the inclination angle. Based on the modelling in the three filters, we choose the best inclination angle value (with the best $\chi^2$ value) for each filter. For the majority of disc-dominated galaxies these three filters give close inclinations as the best values.

In the final modelling step, the $gri$ filter data were used together with the best guess for the inclination angle. The final modelling gives us the structural parameters of galaxy components and the position and inclination angles of the galaxy -- the orientation of its spin vector. The structural parameters together with the morphological classification were used to select mainly disc-dominated or elliptical/S0 galaxies for the correlation study.

Figure~\ref{fig:cosidist} shows the distribution of the inclination angles of galaxies for all  galaxies and for spiral galaxies. Compared with the uniform distribution the number of both face-on and edge-on galaxies is smaller. This is caused by two reasons: firstly, the inclination angle is degenerate with the thickness of the disc, so it is hard to find exactly edge-on galaxies automatically; secondly, real galaxies are triaxial ellipsoids and even for exactly face-on galaxies, the model sees it as a slightly inclined galaxy. In the distribution, we also see the peaks of the initial inclination angle values (ten values from $0\degr$ to $90\degr$). However, the deviation from the random distribution is not severe and it does not bias our correlation studies, it just adds some noise.

If we look at the distribution of the inclination angles of spiral galaxies (Fig.~\ref{fig:cosidist}), we see that these galaxies are mostly edge-on. This is purely a selection effect: spiral galaxies are much easier to classify if they are edge-on. In our spiral galaxy sample, we miss many face-on spiral galaxies. In the correlation study we will take this selection effect into account and it does not influence our results. 

In order to obtain a more reliable detection of the spin orientation angles for the galaxies, the solution of the galaxy model can be improved by using some extra priors or regularisation conditions. We are working in this direction.

\subsection{The sample of spiral and elliptical/S0 galaxies}
\label{sect:galsample}

\begin{figure}
    \includegraphics[width=84mm]{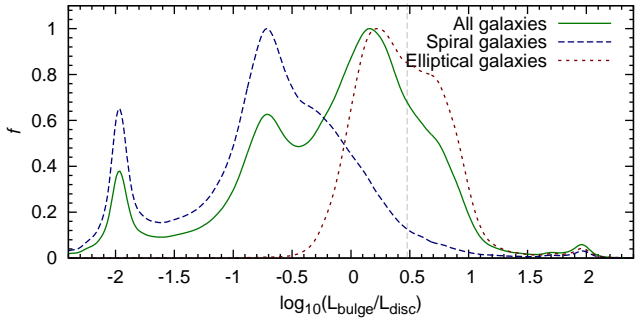}
    \caption{The distribution of the bulge-to-disc ratios. The green solid line -- all galaxies; the blue dashed line -- spiral galaxies; the red dotted line -- elliptical galaxies. The grey vertical dashed line shows the limit $L_\mathrm{bulge}/L_\mathrm{disc}=3$ that we used to select galaxies for the correlation study.}
     \label{fig:btdist}
\end{figure}

To study the correlation between galaxies and filaments we use spiral and elliptical/S0 galaxies. The reason is that the inclination angles of galaxies can be adequately restored only for spiral/S0 galaxies; for purely elliptical galaxies the inclination angles cannot be uniquely recovered. Our morphological classification of galaxies was initially done in \citet{Tempel:11a}. In \citet{Tempel:12} we combined this classification with the morphological classification by the Galaxy Zoo project \citep{Huertas-Company:11} and improved our classification.

In this paper, we use the results of our galaxy modelling and restrict our galaxy sample to  the objects with the bulge-to-disc ratio less than three. Figure~\ref{fig:btdist} shows the distribution of the bulge-to-disc ratios for all, spiral, and elliptical galaxies. Most spiral galaxies have the bulge-to-disc ratio less than two, and most ellipticals have the bulge-to-disc ratio greater than one.

To study the orientation correlation between galaxies and filaments we use two samples. Firstly, we use a spiral galaxy sample with $L_\mathrm{bulge}/L_\mathrm{disc}<2$. The second sample is an elliptical/S0 galaxy sample (by our morphological classification): for this sample we also use only the galaxies with $L_\mathrm{bulge}/L_\mathrm{disc}<3$, representing mostly lenticular galaxies. We note that our morphological classification does not discriminate  between elliptical and S0 galaxies, therefore we use the term elliptical/S0 for this sample.

As a final step, we eliminate from our sample these galaxies which are close to the borders of the survey since we cannot find reliable filament orientation there. Since the minimum filament length in our study is 5\,$h^{-1}$Mpc (see Sect.~\ref{sect:fil}), we set the limiting distance from the survey border to half of it (2.5\,$h^{-1}$Mpc). We use the spatial sample mask as defined in \citet{Martinez:09}.

As a result, our final sample that we use for the correlation study includes 231544 spiral galaxies and 26140 elliptical/S0 galaxies.

\section{Finding the filamentary structure}
\label{sect:fil}

\begin{figure}
    \includegraphics[width=84mm]{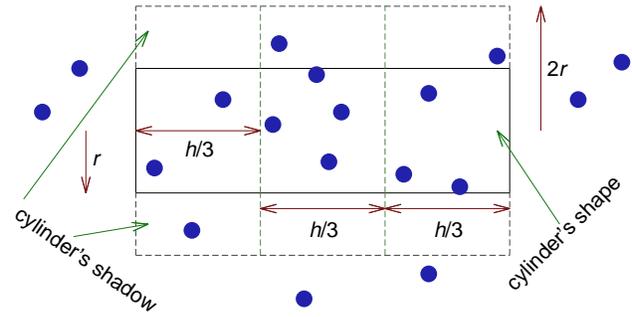}
    \caption{Two dimensional projection of a thin cylinder with its shadow within a pattern of galaxies.}
     \label{fig:cylinder}
\end{figure}

\subsection{Detecting filaments and their orientation}

\begin{figure*}
    \includegraphics[width=180mm]{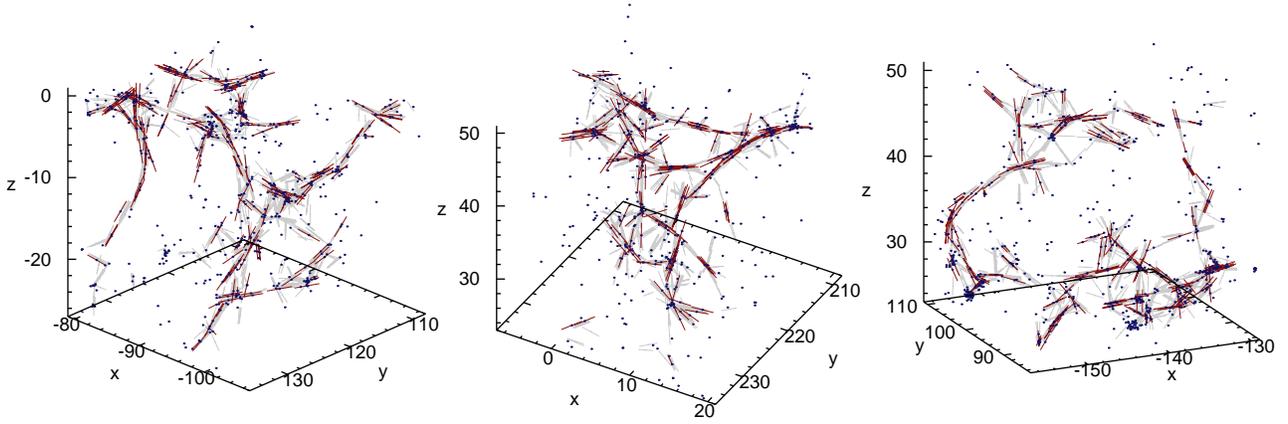}
    \caption{A sample of filaments and their orientations. Three various regions from the SDSS volume (units are in $h^{-1}$Mpc) are shown. Blue dots show the location of galaxies. Grey lines show the location of probed cylinders (see text for more information). Red lines around galaxies show the filaments and their orientation. The figures are rotated so that the longest filaments are well seen.}
     \label{fig:fils}
\end{figure*}

In this section we describe the main tools we use to study the large-scale filaments orientations. Our previous papers \citep{Stoica:07, Stoica:10} concentrate on detecting a general filamentary pattern. In this paper, our main motivation is to detect the filament orientation. To do that, we modify/redefine the underlying model and use this model to detect the orientation of filaments at galaxy locations. This approach is close to that of \citet{Hill:11} who show how to estimate the field of orientations from the point pattern data.

\subsubsection{Elementary cylinder}
\label{sect:elemcyl}

We model the filamentary structure by a network of thin connected cylinders. We assume that locally, galaxies may be grouped together inside a rather small thin cylinder. We also assume that such small cylinders may combine to extend a filament if neighbouring cylinders are aligned in similar directions. We used such a cylinder network to build filaments in our earlier papers \citep{Stoica:07,Stoica:10}. In these papers the cylinders were used to build an object point process that delineated filaments; here we filter the point process given by galaxy positions by the cylindrical kernel to estimate the orientation field. This is similar to the work done in \citet{Aragon-Calvo:07a, Aragon-Calvo:10}, \citet{Sousbie:08}, and \citet{Smith:12}, where the starting point is to determine a smoothed density field and to use its structure to define filaments. The main difference with our approach is that the kernel used to detect filaments in these papers is, by necessity, isotropic, while our kernel is anisotropic, cylindrical, and follows the local anisotropy of the galaxy distribution.

A cylinder $y$ is an object characterised by its centre $k$ and its shape parameters. The shape parameters of a cylinder are the radius $r$, the height $h$, and the orientation vector $\omega$. We consider the radius and the height parameters as fixed, selecting $r=0.5\,{h}^{-1}$Mpc, and $h=5.0\,{h}^{-1}$Mpc. These values fit well the observed filamentary structure. The orientation vector parameters $\omega=\phi(\eta,\tau)$ are uniformly distributed on $M=[0,2\pi]\times[0,1]$, such that
\begin{equation}
    \omega=\left(\sqrt{1-\tau^2}\cos(\eta),\,\sqrt{1-\tau^2}\sin(\eta),\,\tau\right).
\end{equation}
We denote the shape of a cylinder by $s(y)=s(k,r,h,\omega)$.

For every elementary cylinder $y$, we introduce a bigger cylinder with exactly the same parameters as the elementary cylinder, except for the radius, which equals $2r$. Let $\tilde{s}(y)$ be the shadow of $s(y)$ obtained by the subtraction of the initial cylinder from the bigger cylinder. The cylinder and its shadow are shown in Fig.~\ref{fig:cylinder}. Then, each cylinder is divided in three equal volumes along its main symmetry axis, and we denote by $s_1(y)$, $s_2(y)$, and $s_3(y)$ their corresponding shapes, and by $\tilde{s}_1(y)$, $\tilde{s}_2(y)$, and $\tilde{s}_3(y)$ their corresponding shadow shapes.

We describe the galaxy data as the point field given by the set of the galaxy positions, $\bmath{d}=\{d_1,d_2,\dots,d_n\}$. To find a cylinder that is allowed by the data, we check a number of conditions. We have found that it is advantageous to use slightly more strict criteria than we used in our previous papers \citep{Stoica:07, Stoica:10}.

First, we demand that the density of galaxies inside $s(y)$ is sufficiently higher than the density of galaxies in $\tilde{s}(y)$, formalised as follows:
\begin{equation}
    n(\bmath{d}\cap s(y))/\nu(s(y)) > 2\cdot n(\bmath{d}\cap \tilde{s}(y))/\nu(\tilde{s}(y)),
\end{equation}
where $n(\bmath{d}\cap s(y))$ and $n(\bmath{d}\cap \tilde{s}(y))$ are the numbers of galaxies covered by the cylinder and its shadow, and $\nu(s(y))$ and $\nu(\tilde{s}(y))$ are the volumes of the cylinder and its shadow, respectively. So, we require that the density in the cylinder is at least two times larger than in its shadow. Additionally, for every cylinder element $s_i(y)$ we require that the density in the element should be larger than in its shadow:
\begin{equation}
    n(\bmath{d}\cap s_i(y))/\nu(s_i(y)) > n(\bmath{d}\cap \tilde{s}_i(y))/\nu(\tilde{s}_i(y)),
\end{equation}
where $i\in (1,2,3)$ denotes the cylinder parts. 

To ensure that the cylinder is approximately uniformly populated, we require
\begin{equation}
    \frac{ \max(n(\bmath{d}\cap s_1(y), n(\bmath{d}\cap s_2(y), n(\bmath{d}\cap s_3(y)) }
    { \min(n(\bmath{d}\cap s_1(y), n(\bmath{d}\cap s_2(y), n(\bmath{d}\cap s_3(y)) } \lid 5.
\end{equation}
This criteria also helps to forbid cylinders partially inside galaxy groups.

If all those criteria are satisfied, then we accept the cylinder. These criteria are empirical priors, chosen by observational experience. The most common observed filaments are filaments that connect groups, and the thickness of such filaments is about 1\,$h^{-1}$Mpc. This is the reason, why we chose $r=0.5\,h^{-1}$Mpc: such filaments should be the sites of galaxy formation.

If the above rules are satisfied, we calculate the weighted number of galaxies inside a cylinder (its richness):
\begin{equation}
    R(s(y))=\sum_{i=1}^n\frac{r-D_i}{r},
\end{equation}
where $r$ denotes the cylinder radius, $D_i$ denotes the distance of a galaxy from the axis of the cylinder, and summation is over all galaxies inside the cylinder ($n=n(\bmath{d}\cap s(y))$). If the rules listed above are not satisfied, we set the richness (the intensity of the orientation field) at the centre of the cylinder to zero.

The richness $R(s(y))$ is used to estimate the goodness of a cylinder: it is determined by the number density inside a cylinder, and weighting helps to find the best orientation for a cylinder.

\subsubsection{Finding the filaments and their orientation}

To study the correlation between galaxies and filaments, we need to know the filament orientation at the location of a galaxy (if the galaxy is in a filament). To define the filaments and their orientations, we build a filament orientation field that has direction and intensity at every point of the sample volume \citep[see, e.g.,][]{Hill:11}. For that we do the following.
\begin{enumerate}
    \item We use a local morphological filter to detect the direction and an intensity of the filament orientation field. This filter is based on the elementary cylinder defined in the previous Section. This cylinder is defined only in the regions (in location and orientation space) that satisfy the criteria given above and this defines the mask for the field; outside the mask, the intensity of the field is zero. The richness $R(s(y))$ defines the intensity of the field. 
    \item In order to apply this filter, we discretise the location space, covering it with cells of size of 1\,$h^{-1}$Mpc. Our test calculations showed that this spacing is small enough to sample the direction field. We assume that only one cylinder (centre of the cylinder) can be located in such a cell, but clearly a cell can be touched by several cylinders.
    \item To estimate the orientation field, we maximise the richness $R(s(y))$ in every cell inside the mask volume; we accept only those cylinders, which satisfy the criteria given in Sect.~\ref{sect:elemcyl}.
    \item The maximisation of $R(s(y))$ is done using a following maximisation procedure: inside each location cell, the centre position of the cylinder together with its corresponding orientation parameters are moved stochastically till its richness is maximised. For that we use multidimensional nested sampling, applying the Multinest tool \citep{Feroz:08, Feroz:09}. This is a parallel algorithm and  maximises the richness  $R(s(y))$ in reasonable time.
\end{enumerate}

Basically, this is cylinder based data filtering and those accepted cylinders define an intensity map for filaments; the cylinder orientations define an orientation map for filaments. Visually, the cylinders are located in beams along a filament. These maps allow to find the number of the cylinders touching a galaxy and this allows to decide whether a galaxy belongs to the filamentary structure or rather to a cluster (crossing points of filaments).

For every galaxy, in order to find the filament orientation in its location, we firstly find all the cylinders that touch (include) the galaxy. If there are no such cylinders, the galaxy is not located in a filament. If at least 2/3 of the touched cylinders have roughly the same orientation (the angle with the mean direction of all cylinders is less than 20$^\circ$), we say that there is a filament: to be able to apply the 2/3 rule, there must be at least three cylinders that touch the galaxy. If there is no such prominent beam, then probably it is just a crossing point of filaments, where the orientation is unclear. We reject those galaxies and do not use them for our correlation studies.

If there is a filament, we use those cylinders that helped us to decide that there is a filament (that have the angles with the mean direction less than 20$^\circ$) and average their directions to define the orientation of their beam (the orientation of the filament at that point). We use the cylinder richness $R(s(y))$ as the weight for averaging. This algorithm gives sensible filament orientations. We use only those galaxies that belong to the filamentary structure to study the correlation between the orientation of galaxies and of the filament orientation field.

These operations, leading to the computation of the filament orientation value at a position of a galaxy, are closely related to morphological filtering \citep[see, e.g.,][]{Serra:82}.

\subsection{An example of filaments}

\begin{figure}
    \includegraphics[width=84mm]{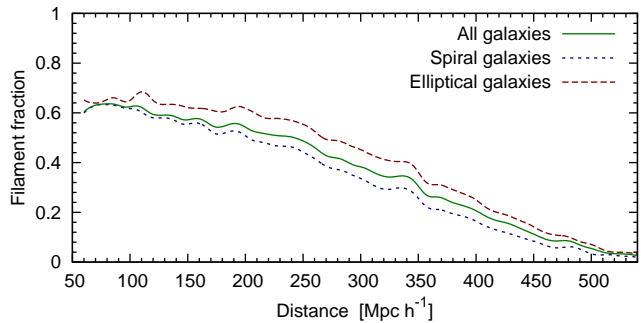}
    \caption{The fraction of galaxies in filaments for all, spiral, and elliptical/S0 galaxies as a function of distance from us.}
     \label{fig:filfrac}
\end{figure}

\begin{figure}
    \includegraphics[width=80mm]{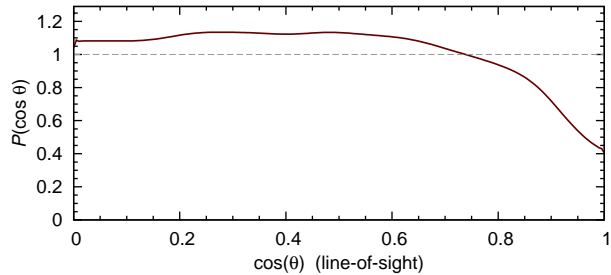}
    \caption{The distribution of the angle between the filaments and the line-of-sight: $\cos{\theta}=1$ for the filaments parallel with the line-of-sight; $\cos{\theta}=0$ for the filaments perpendicular to the line-of-sight.}
     \label{fig:fillos}
\end{figure}

Figure~\ref{fig:fils} shows an example of filaments in three different parts of the SDSS volume. The cubes are rotated so that the best filaments are well seen. We see that the algorithm finds good filaments and the filament orientation is also plausible. A closer look reveals that filaments intersect in the location of groups.

Grey lines in Fig.~\ref{fig:fils} show the allowed cylinders and dark red lines around galaxies mark the galaxies which are in filaments, together with filament orientation.
Initially there are a few cylinders that do not belong to filaments (that are single and that are discarded when calculating filament orientation), but if we test the galaxies against the cylinder pattern, we can eliminate those cylinders (it is well seen in upper right part in central panel of Fig.~\ref{fig:fils}). Our method to find the filaments depends on the number density of objects: if the number density is too low, many of the filaments are undetectable. Figure~\ref{fig:filfrac} shows the fraction of galaxies in filaments as a function of distance from the observer. It is seen that up to 200\,$h^{-1}$Mpc, the fraction is almost constant and decreases afterwards. The fraction of galaxies in filaments is almost the same for spiral and elliptical/S0 galaxies, being slightly smaller for spirals.

Figure~\ref{fig:fillos} shows the distribution of the angles between the filaments and the lines-of-sight. It is seen that our filament finder finds less filaments parallel to the line-of-sight. The reason for this is unclear, but there are two reasonable possibilities. Firstly, since we suppress the finger-of-god redshift distortions, we may over-suppress the groups; or we may include galaxies in the front or back of the groups that actually belong to filaments, into groups. The second reason lies in large-scale cosmological redshift distortions that compress superclusters along the line-of-sight in velocity space \citep[see, e.g.,][]{Praton:1997}. For correlation study, we take this selection effect into account.

\section{Estimating the galaxy-filament correlation}
\label{sect:cor}

To quantify the impact of the filament environment on galaxies, we start with our estimate of the filament orientation field, found using all galaxies in the sample. Using this, we study how the spin vectors of different galaxy subsamples are correlated with the orientation of the filament in which the galaxies reside.

\subsection{Computation of the alignment signal}
\label{sect:comp_align}

To estimate the orientation of galaxies relative to filaments, we compute the empirical probability distribution function $P(\cos{\theta})$ (alignment signal), where $\theta$ is the angle between the rotation axis of a galaxy and the spine/orientation of the filament. The quantity $\cos{\theta}$ is obtained as a scalar product between the unit vectors of spin axes of galaxies ($\mathbf{s}_\mathrm{galaxy}$) and orientation vectors of filaments ($\mathbf{r}_\mathrm{filament}$)
\begin{equation}
    \cos{\theta}=|\mathbf{s}_\mathrm{galaxy}\cdot\mathbf{r}_\mathrm{filament}| .
\end{equation}
Here, $\cos{\theta}$ is restricted to the range $[0,1]$, and $\cos{\theta}=1$ implies that the rotation axis of galaxy is parallel to the filament while $\cos{\theta}=0$ indicates the perpendicular orientation.

The probability distribution function should be compared with the null hypothesis of random mutual orientation of galaxies and filaments. We have seen that neither our galaxies nor filaments have random orientations due to selection effects (see Figs.~\ref{fig:cosidist} and \ref{fig:fillos}). We use Monte-Carlo approximation to estimate the orientation distribution for the case where there are no correlations between the orientations of filaments and galaxies, and to find the confidence intervals for this estimate \citep[see, e.g.,][]{Illian:08}.
For that, we generate 1000 randomised samples in which the orientations (inclination and position angles) of galaxies are kept fixed, but galaxy locations are randomly changed between each other. This gives a new (random) mutual orientation angle between the spin vector of the galaxy and a filament. For each of our subsamples we find the median of the alignment distribution for the null hypothesis together with its 95\% confidence limits. This distribution takes into account both the observational selection effects and the influence of the sample size on the estimate of the distribution. These data are shown in every subsequent figure (Figs.~\ref{fig:corrspirals}--\ref{fig:correllipticalsrnd}), for comparison with the observational results.

To calculate the correlation signal (probability distribution), we use kernel smoothing with a Gaussian kernel of $\sigma=0.045$.

\subsection{Spin vectors of galaxies}

To compute the angle $\theta$ we need to define the spin vector of a galaxy ($\mathbf{s}_\mathrm{galaxy}$) and the orientation vector for a filament ($\mathbf{r}_\mathrm{filament}$). The last one is uniquely defined as described in Sect.~\ref{sect:fil}. For galaxies, we need to know the position angle (orientation in the sky) of a galaxy and its inclination angle. Both of them are estimated by galaxy modelling as described in Sect.~\ref{sect:model}, but since we do not know which side of the galaxy is closer to us, we have two spin vectors for each galaxy (actually four, but we are neglecting the direction of the spin).

Different approaches have been used to deal with the indeterminacy of the spin vectors of galaxies. Since for edge-on and face-on galaxies, the indeterminacy is irrelevant, several studies have used only those galaxies \citep{Trujillo:06, Slosar:09a}. The disadvantage of this approach is the fact that the number of galaxies in the sample is greatly reduced. Another possibility is to use just one spin vector \citep{Lee:07} or both simultaneously \citep{Kashikawa:92, Varela:12}. However, this approach dilutes the strength of the measured signal.

Since we use the statistical approach to calculate the alignment signal, we decided to use both spin vectors of galaxies. \citet{Varela:12} also tested this approach with several Monte Carlo simulations and showed that the procedure recovers correctly the probability distribution function.

\section{Results and discussion}
\label{sect:results}

In Table~\ref{tab:results} we show the number of galaxies, the average of $\cos{\theta}$, and Kolmogorov-Smirnov (KS) test probability that there is no correlation between galaxies and filaments for subsamples presented in this Section. For the KS test, we compared the observed distribution of mutual orientations with the mean of the randomised distributions of mutual orientations ($\bar{P}(\cos\theta)_\mathrm{rnd}$), as explained in Sect.~\ref{sect:comp_align}. In order to compare our results with previous work, we also calculated the average $\langle\cos\theta\rangle$ of the orientation distribution. As the probability density for observed angles $P(\cos\theta)_\mathrm{obs}$ is conditional on the distribution of random orientations $\bar{P}(\cos\theta)_\mathrm{rnd}$,
\begin{equation}
     P(\cos\theta)_\mathrm{obs}=f(\cos\theta)\bar{P}(\cos\theta)_\mathrm{rnd},
\end{equation}
we find the estimate $\hat{f}(\cos\theta)$ from the above formula and use it to estimate the mean $\langle\cos\theta\rangle$.

\begin{table}
 \caption{Number of galaxies, average $\cos{\theta}$, and the KS test probability that the sample is drawn from a uniform mutual orientation for our subsamples of galaxies.}
 \label{tab:results}
 \begin{tabular}{@{}lrccc}
  \hline
  Sample & $N_\mathrm{gal}$ & $\langle\cos\theta\rangle$ & $p_\mathrm{KS}$ & Figure \\
  \hline
  Spirals: all & 67786 & 0.500 & $6.0\cdot 10^{-1}$ & Fig.~\ref{fig:corrspirals} \\
  Spirals: bright & 6640 & 0.508 & $3.3\cdot 10^{-2}$ & Fig.~\ref{fig:corrspirals} \\
  Spirals: test$^{(a)}$ & 6640 & 0.507 & $6.1\cdot 10^{-2}$ & Fig.~\ref{fig:corrspiralseq} \\
  Elliptical/S0: all & 20591 & 0.489 & $3.6\cdot 10^{-7}$ & Fig.~\ref{fig:correllipticals} \\
  Elliptical/S0: bright & 11156 & 0.482 & $1.5\cdot 10^{-9}$ & Fig.~\ref{fig:correllipticals} \\
  Elliptical/S0: inc.$^{(b)}$ & 11156 & 0.482 & $7.6\cdot 10^{-9}$ & Fig.~\ref{fig:correllipticalsrnd} \\
  Elliptical/S0: pos.$^{(c)}$ & 11156 & 0.491 & $9.1\cdot 10^{-4}$ & Fig.~\ref{fig:correllipticalsrnd} \\
  \hline
 \end{tabular}
 
 \medskip
 $^{(a)}$ Refers to the spiral sample, where the inclination is calculated from Eq.~(\ref{eq:inc}).
 $^{(b)}$ Refers to the elliptical/S0 sample where the inclination angles are randomised (see text).
 $^{(c)}$ Refers to the elliptical/S0 sample where the position angles are randomised (see text). 
\end{table}

\subsection{Spiral galaxies}

\begin{figure}
    \includegraphics[width=84mm]{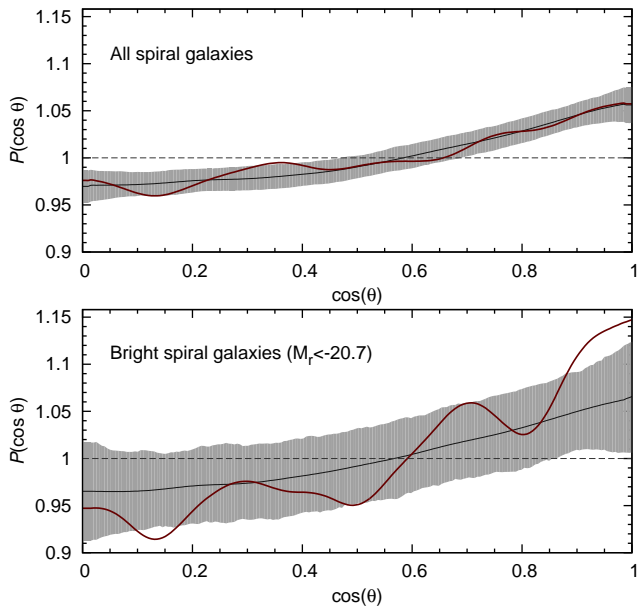}
    \caption{The orientation probability distribution for all (upper panel) and bright (lower panel) spiral galaxies. The black line and the grey filled region show the null hypothesis together with its 95\% confidence limits. The solid red line shows the true alignment distribution.}
     \label{fig:corrspirals}
\end{figure}

\begin{figure}
    \includegraphics[width=84mm]{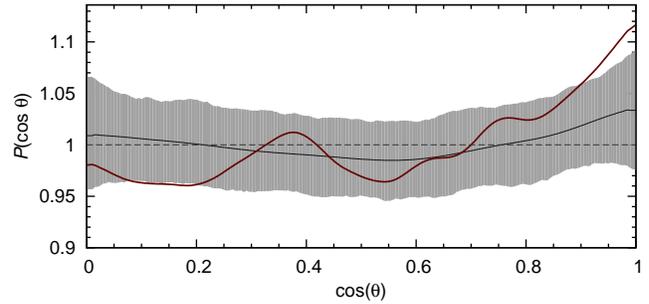}
    \caption{The orientation probability distribution for bright spiral galaxies, where the inclination angles are estimated using Eq.~(\ref{eq:inc}). The black line and the grey filled region show the null hypothesis together with its 95\% confidence limits. The solid red line shows the true alignment distribution.}
     \label{fig:corrspiralseq}
\end{figure}

Figure~\ref{fig:corrspirals} shows the probability distribution $P(\cos{\theta})$ for the mutual angle $\theta$ between the spin axes of spiral galaxies and the orientation vectors of filaments. The grey shaded area corresponds to the random distribution (95\% confidence region) and the dark red line shows the real signal. It is seen that for all spiral galaxies, the real signal lies within the 95\% confidence region for the random distribution.
The KS test also confirms that the observed signal does not differ from the random distribution.

The lower panel in Fig.~\ref{fig:corrspirals} shows $P(\cos{\theta})$ for bright spiral galaxies ($M_r<-20.7\,$mag). For bright spiral galaxies, there are clearly more galaxies which have the spin axes parallel to filaments.  
This is confirmed by the KS test and $\langle\cos\theta\rangle$ (see Table~\ref{tab:results}).
However, since we can model brighter galaxies better (and can better estimate the inclination angles), we cannot definitely say that the correlation is intrinsically stronger for brighter galaxies.

Many studies \citep[e.g.][]{Slosar:09a, Varela:12} use the apparent axial ratio ($b/a$) and an assumed disc flatness ($f$) to restore the inclination angles $i$ of galaxies. They use the formula
\begin{equation}
    \cos^2{i} = \frac{(b/a)^2-f^2}{1-f^2} .
    \label{eq:inc}
\end{equation}
For values of $b/a<f$, the angle is set to $90\degr$. The flatness $f$ depends on the morphological type of galaxies, but usually an average value is used.

To test how this simplified equation can be used to study the correlation between the orientation of galaxies and filaments, we calculated the inclination angles using Eq.~(\ref{eq:inc}) and assuming constant disc flatness $f=0.14$ \citep{Varela:12}, which is in good agreement with the average value from our 3D modelling.

Figure~\ref{fig:corrspiralseq} shows the correlation between the spin axes of spiral galaxies and filaments, when the inclination angles were estimated using Eq.~(\ref{eq:inc}). It is seen that the correlation is weaker, showing that the simplified equation is not a good tool to estimate the inclination angles of galaxies. As a conclusion, detailed photometric modelling is necessary to restore the inclination angles accurately.

Since the correlation signals in Figs.~\ref{fig:corrspirals} and \ref{fig:corrspiralseq} are practically the same, it shows that the correlation signal is real and is unaffected by the method to estimate the inclination angles of galaxies (e.g. the peaky distribution in Fig.~\ref{fig:cosidist}).

\subsection{Elliptical/S0 galaxies}

\begin{figure}
    \includegraphics[width=84mm]{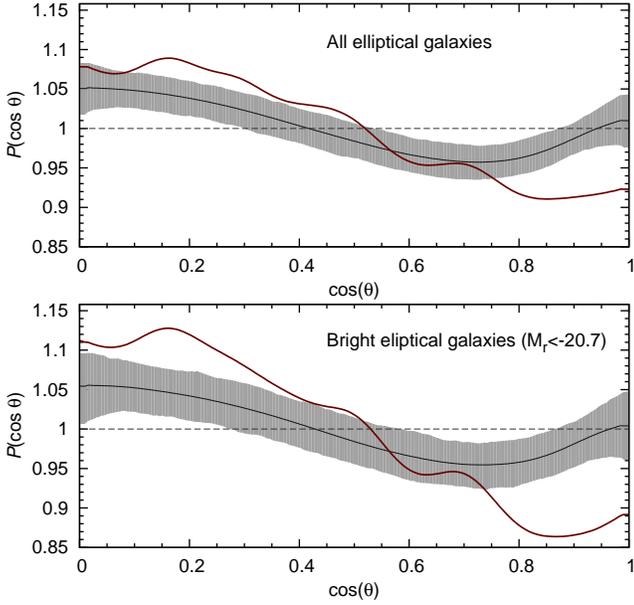}
    \caption{The orientation probability distribution for all (upper panel) and bright (lower panel) elliptical/S0 galaxies. The black line and the grey filled region show the null hypothesis together with its 95\% confidence limit. The solid red line shows the true alignment distribution.}
     \label{fig:correllipticals}
\end{figure}

\begin{figure}
    \includegraphics[width=84mm]{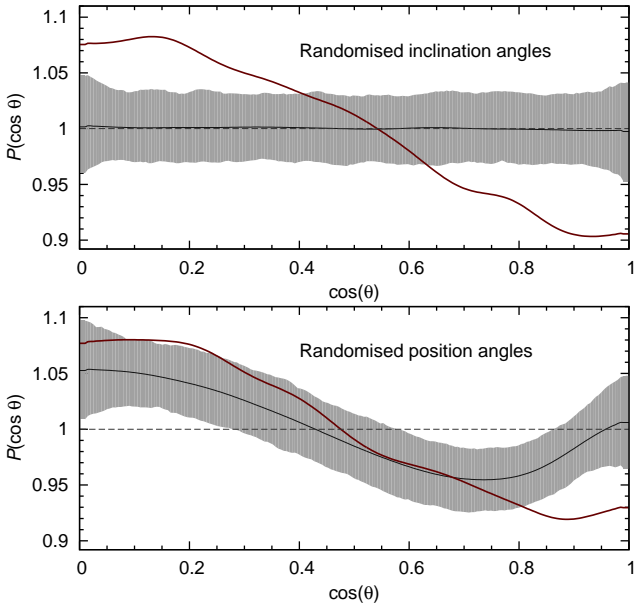}
    \caption{The orientation probability distribution for bright elliptical/S0 galaxies. The upper panel shows the distribution when inclination angles are randomised. The lower panel is for randomised position angles.}
     \label{fig:correllipticalsrnd}
\end{figure}

We also study the correlation for elliptical/S0 galaxies. For that we restrict our sample to galaxies where the bulge-to-disc ratio is less than three, because for higher bulge-to-disc ratios, the inclination angle cannot be restored.

Figure~\ref{fig:correllipticals} shows the $\cos{\theta}$ probability distribution for all and for bright elliptical/S0 galaxies. As for spiral galaxies, the correlation signal is larger for brighter galaxies. 
(The KS test $p$-values are $3.6\cdot10^{-7}$ and $1.5\cdot10^{-9}$, respectively.)

Actually, if we use only faint elliptical/S0 galaxies, the signal is almost lost. There might be two reasons: firstly, the correlation signal is weaker for fainter galaxies; secondly, this is purely a selection effect. For fainter galaxies, the morphological classification is less accurate and also the PSF makes galaxies rounder and the position/inclination angles are much harder to estimate. Compared with spiral galaxies, the correlation signal is much stronger. We note that the possible formation mechanism for S0 and elliptical galaxies may be different (formers are rotationally supported) and the resulting signal is mixed. However, our sample should be dominated by elliptical galaxies and leaving out S0 galaxies, the effect we see in the correlation of ellipticals should be even stronger.

Since the inclination angles for elliptical/S0 galaxies cannot be restored with high accuracy, we studied how the correlation signal depends on inclination angles. For this tests, we used the bright elliptical/S0 sample. In the first test, we randomised the inclination angles of galaxies. The results are shown in the upper panel of Fig.~\ref{fig:correllipticalsrnd}. It is seen that for randomised inclination angles, the correlation signal remains, meaning that the signal comes from filaments which are perpendicular to the line-of-sight. If the spin axis of a galaxy is perpendicular to such a filament, changing the inclination angle does not change the angle between the spin of the galaxy and the filament.

In the second test, we kept the inclination angles fixed and randomised the position angles of galaxies. The results are shown in the lower panel of Fig.~\ref{fig:correllipticalsrnd}. The correlation signal is much weaker in this case, indicating that the inclination angles are not accurately restored. However, the correlation signal is still visible (the KS test $p$-value is $9.1\cdot10^{-4}$), meaning that the modelled inclination angles are statistically close to the real ones.

As a conclusion, the correlation signals for elliptical/S0 galaxies can also be calculated, regardless of the inaccuracy in the determination of the inclination angle. Moreover, the correlation signal is so strong, that it is visible even if the inclination angles are randomised.

\subsection{Comparison with $N$-body simulations and previous observational detections}

Many $N$-body simulations indicate that the correlation of spin axes of dark matter haloes with filaments (sheets) is halo mass dependent \citep{Aragon-Calvo:07, Zhang:09, Trowland:12, Codis:12}: spin axes of low-mass haloes tend to lie parallel to the filaments while high-mass haloes have an orthogonal alignment. To be more specific, \citet{Hahn:10} studied especially the orientation of disc galaxies in a hydrodynamic AMR simulation. They showed that the most massive galaxy discs have spins preferentially aligned so as to point along their host filaments. The same conclusion was reached in \citet{Libeskind:12}, who claimed that low-mass haloes (that mostly host spirals) tend to be aligned with filaments. Both papers are in good agreement with our study, where we found that especially for brighter disc galaxies, the galaxy spins tend to be aligned parallel to the filaments.

To compare $N$-body simulations with our findings, we link low-mass haloes with spiral galaxies and high-mass haloes with elliptical galaxies. The probable formation mechanisms for low- and high-mass haloes support these connections. The spin directions of low-mass haloes  result from the winding flows from the sheets into filaments. Such a process naturally produces a net halo spin parallel to the filaments. Since those haloes are not influenced by many mergers, they are natural hosts for spiral galaxies.

On the other hand, high-mass haloes are significantly affected by mergers occurring mostly along the filaments. Therefore they acquire a spin which is preferentially perpendicular to filaments. These merger-produced haloes host also elliptical galaxies which form through many mergers.

We note that the above correspondence between galaxies (spirals and ellipticals) and haloes (low and high mass) is oversimplified, but considering the fact that derived correlation signal is relatively weak, this simplified picture might explain the difference between those two correlation signals. Moreover, the weakness of the correlation signal can be also explained by the fact that $N$-body simulations show only a weak trend of merger rate with halo mass, weakening the correlation expected from the simplified approach.

Contrary to $N$-body simulations, where there are many papers devoted to the study of the correlation between the orientation of galaxies and their host structures, there is little observational evidence for that. The reasons for that are, however, clear: it is hard to estimate the spin axes of galaxies and to identify the structures where galaxies are located (due to observational effects). Regardless of these difficulties, there are still some papers devoted to this problem. Recently, \citet{Jones:10} selected 69 well-defined edge-on galaxies and used the Multi-scale Morphology Filter (MMF) technique to detect filaments. They found that spin axes of less intrinsically bright galaxies tend to lie perpendicular to host filaments. Based on the study of voids \citet{Varela:12} found that spin axes of disc galaxies around the voids are preferentially directed toward the centres of the voids. Since filaments are located around the voids, the spin axes are also perpendicular to filaments.

Both these findings contradict with our results. However, our present work agrees with \citet{Trujillo:06} and \citet{Lee:07}, where the authors found that spin axes of spiral galaxies are preferentially in the plane of the walls that delineate the cosmic voids and parallel to filaments. In short, the observational findings seem to be in disagreement with each other and the results remain inconclusive. The reasons for that are still unclear. The biggest problem with observational findings is the statistical significance of the results and until this improves, we cannot say anything convincing.

Our results support the tidal torque scenario where the galaxy spins indeed originated from the initial tidal interaction with the surrounding matter. Moreover, our results are supported by  recent high-resolution $N$-body simulations \citep{Trowland:12, Codis:12}.

\section{Concluding remarks}

We have examined the orientation of spin axes of galaxies with respect to the cosmic  filaments. We used 3D photometric modelling of galaxies to restore the inclination angles of galaxies with a sufficiently high accuracy, especially for disc-dominated galaxies. The alignment between galaxy spins and the axis of filaments was characterised by the shape of the probability distribution of $\cos{\theta}$ where $\theta$ is the angle between the two vectors. We studied the correlation for spiral and elliptical/S0 galaxies separately.

We found a strong and significant correlation between the spin axes of elliptical/S0 galaxies and filaments; these galaxies tend to be aligned perpendicular to filaments, whereas more luminous elliptical/S0 galaxies have a stronger orthogonal alignment. We showed that this finding is not influenced by the fact that the inclination angles for elliptical/S0 galaxies cannot be estimated accurately.

Within the caveats of a limited statistical significance, we found that the spin axes of 
bright spiral galaxies tend to be aligned parallel to the host filaments. The significant  alignment we have found is due to finding the inclination angles of galaxies by 3D modelling: using a simple inclination angle estimate by the visible axial ratio produces a weaker correlation. This emphasises the benefits of having a good estimate for inclination angles when studying the correlation of galaxies with the cosmic web.

Our findings for spiral and elliptical/S0 galaxies are in agreement with the recent results of $N$-body simulations, suggesting that spirals form through peaceful accretion of matter and ellipticals are the results of mergers that mostly occur along the filaments.

We hope that our results provide a better understanding of the formation of galaxies in the context of their host large-scale structures, which is one of key questions for galaxy formation theory. A precise knowledge of correlated galaxy orientations is of high importance also for future high-precision weak lensing studies of dark energy. So, it is of high importance to improve the filament finding algorithms and the 3D modelling of galaxies to have a better observational insight into this problem. Future work is planned in this direction.

\section*{Acknowledgments}

We would like to acknowledge helpful discussions with A.~Hirv and L.~J.~Liivam\"agi. The work was greatly inspired by discussions with Rien van~de~Weijgaert. We thank the Leibniz-Institut f\"ur Astrophysic Potsdam, where a vital part of this study was performed, for hospitality. We thank our referee for questions and suggestions that helped to improve the presentation. We acknowledge the support by the Estonian Science Foundation grants 7765, 8005, 9428, MJD272; the Estonian Ministry for Education and Science research project SF0060067s08; the Centre of Excellence of Dark Matter in (Astro)particle Physics and Cosmology (TK120). All the figures have been made using the gnuplot plotting utility. Most of the time-consuming computations were carried out at the High Performance Computing Centre, University of Tartu.

We are pleased to thank the SDSS-III Team for the publicly available data releases. Funding for the SDSS-III has been provided by the Alfred P. Sloan Foundation, the Participating Institutions, the National Science Foundation, and the U.S. Department of Energy Office of Science. The SDSS-III web site is \texttt{http://www.sdss3.org/}.


\label{lastpage}

\end{document}